\documentclass[12pt,a4paper,]{article}
\PassOptionsToPackage{unicode}{hyperref}
\PassOptionsToPackage{naturalnames}{hyperref}
\usepackage{amsmath}
\usepackage{amssymb}
\usepackage[hidelinks]{hyperref}
\usepackage{multirow}
\usepackage{float}
\usepackage{array}
\usepackage[numbers,sort&compress]{natbib}
\usepackage{mathtools}
\usepackage{graphicx}
\usepackage{subcaption}
\usepackage{makeidx}
\usepackage[version=4]{mhchem}
\usepackage{lipsum}
\usepackage{footmisc}
\usepackage{microtype}
\usepackage{pgfplots}
\usepackage{xcolor}
\usepackage{slashed}
\usepackage[table]{xcolor}
\usepackage[most]{tcolorbox}
\usepackage{titlesec} 
\titleformat{\subparagraph}
{\normalfont\normalsize\bfseries}{\thesubparagraph}{1em}{}
\titlespacing*{\subparagraph}
{1.5em}{2.5ex plus 1ex minus .2ex}{1.0ex plus .2ex}
\tcbset{
	colback=yellow!20,
	colframe=yellow!50!black,
	boxrule=0pt,
	left=4pt,right=4pt,top=4pt,bottom=4pt
}

\pgfplotsset{compat=1.18}

\title{\textsl{Rare Exclusive Top Decays $t\to bM$ and Vector-Meson Helicity}}

\author{
	\large M.Ahmadi\textsuperscript{1}, M.Monemzadeh\textsuperscript{2}, N.Tazimi\textsuperscript{3} \\
	\small \emph {Department of Physics, University of Kashan, Kashan, Iran}}
\date{}
\begin{document}
	\maketitle
	\sloppy
	\vspace{.9cm}
	\footnotetext[1]{{\em E-mail address:} maryam.ahmadi@grad.kashanu.ac.ir}
	\footnotetext[2]{{\em E-mail address:} monem@kashanu.ac.ir}
	\footnotetext[3]{{\em E-mail address:} tazimi@kashanu.ac.ir}

\begin{abstract}
We present a leading-order study of the rare exclusive decays $t\to bM$, where $M$ denotes a pseudoscalar or vector meson. Starting from the tree-level transition $t\to bW^\ast\to b(q\bar q')$ and using a factorization approximation, we derive compact expressions for the amplitudes and partial widths in which the long-distance QCD dynamics is encoded in the meson decay constants $f_M$. We provide numerical branching-ratio predictions for representative modes such as $t\to b\pi^+$, $t\to bK^+$, $t\to bD_s^{(\ast)}$ and $t\to bB_c^{(\ast)}$, finding $\mathrm{Br}\sim 10^{-8}$-$10^{-7}$. For vector final states we derive helicity amplitudes and Standard-Model polarization fractions, and we outline the sensitivity of these observables to anomalous chiral $tWb$ couplings in the limit $m_b\to 0$. Our results provide an updated Standard Model (SM) baseline for rare exclusive top-quark decays and a starting point for EFT interpretations of non-standard charged-current interactions.
\end{abstract}

\section{Introduction}

The top quark is the heaviest particle in the Standard Model (SM) and provides a unique laboratory for precision studies of the charged–current interaction and for indirect searches of physics beyond the SM. Due to its large mass, the top quark decays before hadronizing, and its dominant decay mode is the transition $t\to bW$, with a total width of order $1.42~\mathrm{GeV}$ and a branching ratio very close to unity. The structure of the tWb vertex is therefore tightly connected to a broad program of top-quark measurements at the LHC, including the determination of the W boson helicity fractions, differential decay distributions, and constraints on effective-field-theory (EFT) operators affecting charged–current interactions.

In addition to inclusive observables, rare exclusive hadronic top decays offer a complementary, theoretically clean set of benchmarks. In this class of processes, the top quark undergoes the transition
\[
t \to b W^{\ast} \to b \, (q\bar q') \to b\,M,
\qquad M=P,V
\]

where M denotes a pseudoscalar (P) or vector (V) meson formed through the hadronization of the light quark pair produced by the off-shell (W) boson. Within the Standard Model (SM), the branching fractions for exclusive top-quark decays into light mesons are predicted to be highly suppressed, typically of order $10^{-8}-10^{-7}$ for CKM-favored channels~\cite{Altarelli}. Despite their tiny branching fractions, these decays provide a well-defined SM benchmark for future high-statistics top-quark studies at the High-Luminosity LHC (HL-LHC) and future colliders. They also offer a clean framework for investigating possible non-standard tWb interactions in exclusive decay modes. In addition, decays into vector mesons provide access to polarization observables, such as helicity fractions and angular distributions, which can be measured experimentally and directly compared with the well-established analysis techniques developed for the standard $t\to bW$ decay.

From a theoretical viewpoint, the exclusive modes $t\to bM$ are particularly transparent. Since the initial state contains a free top quark (rather than a bound hadron), the short-distance transition $t\to bW$ can be described at leading order by quark spinors and the charged–current interaction, without introducing nonperturbative heavy-to-heavy form factors. In the factorization approximation, all long-distance QCD dynamics associated with the formation of the meson $M$ is encoded in a single decay constant ($f_M$). Non-factorizable contributions and QCD radiative corrections are expected to be suppressed by $\alpha_s(m_t)$ and by power corrections in $\Lambda_{\mathrm{QCD}}/m_t$, and can therefore be treated as controlled theoretical uncertainties at the present level of precision.

In this work we present a leading-order analysis of the exclusive decays $t\to bM$ with $M=P,V$. Starting from the full tree-level amplitude $t\to bW^{\ast}\to b \, (q\bar q')$, we derive compact factorized expressions for the decay amplitudes and partial widths, retaining the exact dependence on the meson mass through the two body phase space. For simplicity, we absorb the common short-distance normalization associated with the off-shell $W^\ast$ propagator into our effective leading-order normalization. For vector final states we include the full polarization sum and provide formulas suitable for a helicity decomposition. We then update SM branching-ratio predictions for representative channels $\pi^+,K^+,D_s^{(\ast)+},B_c^{(\ast)+}$, quantify parametric and theoretical uncertainties, and perform a set of analytic and numerical cross-checks against limiting cases and previous literature. Our results provide a self-contained and reproducible SM baseline that can be straightforwardly extended to include anomalous tWb couplings in an EFT framework.

In Sec.~\ref{sec:theory} we derive the factorized amplitudes and decay-width formulas for the exclusive decays $t\to bP$ and $t\to bV$, and specify our normalization conventions. In Sec.~\ref{sec:results} we present numerical predictions using up-to-date input parameters, estimate the main parametric and theoretical uncertainties, and compare with previous studies. Helicity amplitudes and polarization fractions for vector final states are discussed in Sec.~\ref{sec:vector-helicity}. Finally, Secs.~\ref{sec:discussion} and \ref{sec:conclusion} contain a discussion of the phenomenological implications, summarize our main results, and outline possible extensions, including helicity-based observables and EFT interpretations.

\section{Theoretical framework}
\label{sec:theory}

In the SM, the dominant decay mode of the top quark is the charged–current transition

\begin{equation}
	t \to b\, W^+
	\label{tb}
\end{equation}

with a branching ratio very close to unity \cite{PDG}. Rare hadronic modes of this transition are the type

\begin{equation}
	t \to b\, M \,, \qquad 	M = P, V  
	\label{tbm}
\end{equation}

where $P$ and $V$ denote pseudoscalar and vector mesons, respectively, proceed via a virtual (off–shell) $W$ boson with invariant mass fixed by the meson,

\begin{equation}
	t(p_t) \to b(p_b) \, W^{\ast}(q) \to b(p_b) \, q(p_q) \, \bar q'(p_{\bar q'}) \to b\, M
	\label{twb-bqq}
\end{equation}

The relevant SM Lagrangian interaction is the charged–current 

\begin{equation}
 	 \mathcal{L}_{\rm CC}  = -\frac{g}{2\sqrt{2}}\, 
 	 \left[\bar{u}_i \gamma^\mu (1-\gamma_5) V_{ij} d_j\, W^+_\mu +\bar{\nu}_\ell \gamma^\mu (1-\gamma_5) \ell\, W^+_\mu \right] + \text{h.c.}
 	 \label{LCC}
\end{equation}

Here, $g$ is the $SU(2)_L$ weak coupling constant, $W_\mu^+$ denotes the charged $W^+$ boson field, and $\gamma^\mu$ and $\gamma_5$ are the Dirac gamma matrices. The fields $u_i$ and $d_j$ represent the initial and final quark fields, respectively. The quantities $V_{ij}$ are the elements of the Cabibbo-Kobayashi-Maskawa (CKM) matrix, which describe the mixing between the quark generations in charged-current weak interactions. The fields $\nu_\ell$ and $\ell$ denote the neutrino and charged-lepton fields, with $\ell=e,\mu,\tau$. The factor $(1-\gamma_5)$ reflects the left-handed chiral structure of the weak charged-current interaction, while $h.c.$ denotes the Hermitian conjugate of the interaction.
	
The tree–level amplitude for transition Eq. \ref{twb-bqq} is given by 

\begin{equation} 
  	\mathcal{A}(t\to b  q\bar q') = \frac{g^2}{8} \frac{1}{q^2 - m_W^2} \left[ \bar u_b(p_b) \gamma^\mu (1-\gamma_5) V_{tb} u_t(p_t) \right] \left[ \bar u_q(p_q) \gamma_\mu (1-\gamma_5) V_{qq'}^\ast v_{q'}(p_{\bar q'}) \right]
  	\label{Amplitudebqq}
\end{equation}
  	
Since the invariant mass carried by the virtual $W$ boson is fixed by the meson mass, 
$q^2 = m_M^2$, and satisfies $m_M^2 \ll m_W^2$ for all channels considered in this work, the $W$-boson propagator can be expanded in powers of $q^2/m_W^2$. Keeping only the leading term yields

\begin{equation}
	\frac{1}{q^2-m_W^2} \simeq -\frac{1}{m_W^2} \left[ 1+\mathcal{O}\!\left(\frac{q^2}{m_W^2}\right) 	\right]
	\label{FermiExpansion}
\end{equation}

Using the standard definition of the Fermi constant, $\frac{G_F}{\sqrt{2}} = \frac{g^2}{8m_W^2}$, the tree-level amplitude can be written in terms of an effective four-fermion interaction

\begin{equation} 
	\mathcal{A}(t\to b  q\bar q') = \frac{G_F}{\sqrt{2}} V_{tb} V_{qq'}^\ast \left[ \bar u_b(p_b) \gamma^\mu (1-\gamma_5) u_t(p_t) \right] \left[ \bar u_q(p_q)\gamma_\mu(1-\gamma_5) v_{q'}(p_{\bar q'}) \right]
	\label{Amplitudebqq2}
\end{equation}

The neglected higher-order terms are suppressed by $\mathcal{O}(q^2/m_W^2)$ and satisfy

\begin{equation}
	\frac{q^2}{m_W^2} = \frac{m_M^2}{m_W^2}	\lesssim 10^{-3},
\end{equation}

even for the heaviest mesons considered, while for light mesons such as the $\pi$ and $K$ the suppression is several orders of magnitude stronger.

In the exclusive decay $t \rightarrow bM$, the meson $M$ is formed from the hadronization of the $q \bar q'$ pair. by using Factorization method in this decay mode, our amplitude can be written as ~\cite{BenekeNeubert:2003}

\begin{equation}
	\mathcal{A}(t\to b M) = \frac{G_F}{\sqrt{2}} V_{tb} V_{qq'}^\ast \langle M | \bar q \gamma_\mu (1-\gamma_5) q' |0\rangle \, \bar u_b(p_b) \gamma^\mu (1-\gamma_5) u_t(p_t) 
	\label{Amplitudetbm}
\end{equation}

The factorization approximation adopted in this work (Eq.~\ref{Amplitudetbm}) is motivated by the large hierarchy between the top-quark mass and the QCD scale, $m_t \gg \Lambda_{\rm QCD}$. Within this leading-order approximation, the hadronic matrix element is written as the product of a meson decay constant and the corresponding short-distance $t \to b$ quark current. Non-factorizable QCD effects, including soft-gluon interactions between the emitted meson and the heavy-quark current, are not calculated explicitly in the present analysis and are therefore regarded as a potential source of theoretical uncertainty. Thus, the factorized treatment should be understood as a leading-order approximation rather than as a complete proof of factorization for the present decay.

A systematic study of non-factorizable corrections would require a more detailed treatment of the relevant QCD dynamics. Such corrections could, in principle, be investigated using appropriate effective-field-theory methods, including Soft-Collinear Effective Theory (SCET).

This picture is analogous in spirit to the QCD factorization framework for non-leptonic $B$ decays~\cite{BenekeNeubert:2003}, but has two important differences:
 
\begin{itemize}
 	\item The typical hard scale is set by $m_t\sim 170~\text{GeV}$, so $\alpha_s(m_t)$ is small and long-distance gluon exchanges are expected to be less important than in $B$ decays.
 	\item The $t\to b$ transition occurs between essentially free quarks, without an initial bound state. As a consequence, no non-perturbative $t\to b$ form factors appear at leading order, and all long-distance dynamics is encoded in the meson decay constant $f_M$.
\end{itemize}
 
\subsection{Calculation of Decay Amplitude and Branching Ratios at \texorpdfstring{$t\to b W^\ast \to b M$}{t→bW\ast→bM}}
\label{sec:sm} 

The decay constant for a pseudoscalar meson $P$ with momentum $p$ can be written as

\begin{equation}
	\langle P(p) |\, \bar q\, \gamma_\mu \gamma_5\, q'\, |0\rangle = i f_P\, p_\mu\,
	\label{decayconstp}
\end{equation}

while for a vector meson $V$ with momentum $p$ and polarization $\varepsilon_\mu$ is

\begin{equation}
	\langle V(p,\varepsilon) |\, \bar q\, \gamma_\mu\, q'\, |0\rangle = f_V\, m_V\, \varepsilon_\mu
	\label{decayconstv}
\end{equation}

The amplitude for $t\to b P$ and $t \to b V$ can be written as

\begin{equation}
	\mathcal{A}(t\to b P) = \frac{G_F}{\sqrt{2}}\, 	V_{tb} V_{qq'}^\ast \, f_P\, p_P^\mu\, \bar u_b(p_b)\,\gamma_\mu (1-\gamma_5)\,u_t(p_t)
	\label{amplitudeqcdp}
\end{equation}

\begin{equation}
	\mathcal{A}(t\to b V) = \frac{G_F}{\sqrt{2}}\, V_{tb} V_{qq'}^\ast \, f_V m_V\, \varepsilon_\mu\, \bar u_b(p_b)\,\gamma^\mu (1-\gamma_5)\,u_t(p_t)
	\label{amplitudeqcdv}
\end{equation}

For any two-body decay $X \to Y_1,Y_2$, the partial width can be written as \cite{Goodsell:2017}

\begin{equation}
	\Gamma(X \to Y_1 Y_2) = \frac{1}{16\pi m_X^3} \lambda^{1/2}(m_X^2,m_{Y_1}^2,m_{Y_2}^2) 	\overline{|\mathcal{A}|^2}
	\label{gamma}
\end{equation}

where X denotes the initial particle and $Y_1$ and $Y_2$ are the final-state particles. The notation $\overline{|\mathcal{A}|^2}$ denotes the squared amplitude summed over the spins and polarizations of the final-state particles and averaged over the spin of the initial particle. For the decay $t\to b,M$, this gives

\begin{equation}
	\Gamma(t\to b M) = \frac{1}{16\pi m_t^3} \lambda^{1/2}(m_t^2,m_b^2,m_M^2) \overline{|\mathcal{A}(t\to b M)|^2}
	\label{gammaf}
\end{equation}

The Källén function $\lambda(x,y,z)$ is defined by

\begin{equation}
	\lambda(x,y,z) = x^2+y^2+z^2-2xy-2xz-2yz
	\label{lambda}
\end{equation}

In the rest frame of the decaying top quark, the magnitude of the three-momentum of either final-state particle is

\begin{equation}
	|\vec p| = \frac{1}{2m_t} \sqrt{\lambda(m_t^2,m_b^2,m_M^2)}
	\label{momentum}
\end{equation}

\paragraph{Pseudoscalar final states}

For pseudoscalar final states, the spin-averaged squared amplitude can be written as

\begin{equation}
	\overline{|\mathcal{A}(t\to bP)|^2} = G_F^2 |V_{tb}V_{qq'}^\ast|^2 f_P^2 \mathcal{K}_P(m_t,m_b,m_P)
\end{equation}

where the kinematic function $\mathcal{K}_P$ is

\begin{equation}
	\mathcal{K}_P(m_t,m_b,m_P) = (m_t^2-m_b^2)^2 - m_P^2(m_t^2+m_b^2)
	\label{KP}
\end{equation}

Combining the spin-averaged squared amplitude with the standard two-body phase space gives

\begin{equation}
	\Gamma(t\to b P) = \frac{G_F^2}{32\pi m_t^3} \, |V_{tb}V_{qq'}^\ast|^2 \,
	f_P^2 \, \lambda^{1/2}(m_t^2,m_b^2,m_P^2) \, \mathcal{K}_P(m_t,m_b,m_P)
	\label{gammap}
\end{equation}

The detailed derivation of the pseudoscalar Dirac trace is presented in Appendix~\ref{app:pseudoscalar}.

\paragraph{Vector final states}

For vector meson final states, the squared amplitude is obtained by summing over the spins of the initial and final quarks and over the three physical polarizations of the vector meson. The factorized amplitude reads

\begin{equation}
	\mathcal{A}(t\to bV_\lambda) =\frac{G_F}{\sqrt2}\,V_{tb}V_{qq'}^\ast\,f_V m_V\,
	\varepsilon_\mu^\ast(p_V,\lambda)\, \bar u_b(p_b)\gamma^\mu(1-\gamma_5)u_t(p_t)
	\label{AV}
\end{equation}

Summing over polarizations

\begin{equation}
	\sum_{\lambda=0,\pm}\varepsilon_\mu^\ast(p_V,\lambda)\varepsilon_\nu(p_V,\lambda) =	-\,g_{\mu\nu}+\frac{p_{V\mu}p_{V\nu}}{m_V^2}
	\label{polsum}
\end{equation}

We obtain

\begin{equation}
	\overline{|\mathcal{A}(t\to bV)|^2} =\frac{G_F^2}{2}\,|V_{tb}V_{qq'}^\ast|^2\,f_V^2m_V^2\,
	L_{\mu\nu}\left(-g^{\mu\nu}+\frac{p_V^\mu p_V^\nu}{m_V^2}\right)
	\label{Vsq-start}
\end{equation}

where the spin average is included and

\begin{equation}
	L_{\mu\nu}=\frac12 \mathrm{Tr}\Big[(\slashed p_b+m_b)\,\gamma_\mu(1-\gamma_5)\,(\slashed p_t+m_t)\,\gamma_\nu(1-\gamma_5)\Big]
	\label{LmunuV}
\end{equation}

Compared to the pseudoscalar case, here the Dirac structure remains V-A and one cannot reduce the amplitude to a single chiral projector by equations of motion.

Evaluating the trace and performing the contraction with the polarization sum one finds the compact result

\begin{equation}
	\overline{|\mathcal{A}(t\to bV)|^2} = G_F^2\,|V_{tb}V_{qq'}^\ast|^2\,f_V^2m_V^2\,  \mathcal{K}_V(m_t,m_b,m_V)
	\label{Vsq-final}
\end{equation}

with

\begin{equation}
	\mathcal{K}_V(m_t,m_b,m_V) 	= \frac{(m_t^2-m_b^2)^2}{m_V^2} +  m_t^2+m_b^2-2m_V^2
	\label{KV-correct}
\end{equation}

The complete evaluation of the vector-meson Dirac traces together with the polarization contraction is given in Appendix~\ref{app:vector}.

Combining this kinematic factor $\mathcal{K}_V$ with the standard two body phase space, the partial width for final vector meson is

\begin{equation}
	\Gamma(t\to b V) = \frac{G_F^2}{32\pi m_t^3}\, |V_{tb} V_{qq'}^\ast|^2\, f_V^2 m_V^2\, \lambda^{1/2}(m_t^2,m_b^2,m_V^2)\, \mathcal{K}_V(m_t,m_b,m_V)
	\label{gammav}
\end{equation}

For pseudoscalar mesons, $P = \pi^+, K^+, D_s, B_c$, the decay widths are obtained from the general expression above using the appropriate CKM factors $V_{qq'}^\ast$ and decay constants $f_P$. For vector mesons $V = \rho^+, K^{\ast+}, D_s^\ast, B_c^\ast$, the analysis proceeds analogously, with the replacement $f_P \to f_V$ and $\mathcal{K}_P \to \mathcal{K}_V$. The branching ratio is then

\begin{equation}
	\text{Br}(t\to b P(V)) = \frac{\Gamma(t\to b P(V))}{\Gamma_{\rm tot}(t)}
	\label{BR}
\end{equation}

With $\Gamma_{\rm tot}(t)$ the total top–quark width. The dominant channels are those enhanced by $|V_{tb} V_{ud}|$ and $|V_{tb} V_{cs}|$, such as $t\to b \pi^+$ and $t\to b D_s$, while modes involving $V_{ts}$ or $V_{td}$ are strongly CKM suppressed and not calculated in this work.

\subsection{Validity of the Factorization Approximation}
\label{sec:factorization}

The exclusive decays considered in this work are treated within the leading-order factorization approximation. In this framework, the short-distance weak transition and the hadronization of the light-quark current into the final-state meson are separated. The decay amplitude
can therefore be expressed as a product of the short-distance electroweak contribution and the vacuum-to-meson matrix element of the corresponding quark current.

As mentioned earlier for a pseudoscalar meson, the relevant matrix element is defined by

\begin{equation}
	\langle P(p_P)|\bar q\gamma^\mu\gamma_5 q'|0\rangle = i f_P p_P^\mu 
\end{equation}

whereas for a vector meson one has

\begin{equation}
	\langle V(p_V,\epsilon)|\bar q\gamma^\mu q'|0\rangle = f_V m_V \epsilon^{*\mu}
\end{equation}

The nonperturbative dynamics associated with the formation of the color-singlet meson is thus encoded in the corresponding decay constant, while the remaining part of the amplitude is determined by
the short-distance weak interaction.

At leading order in QCD, the weak decay proceeds through the color singlet charged-current interaction mediated by the virtual $W$ boson. Consequently, no hard gluon exchange is required to produce the
factorized leading-order amplitude. The factorized description is therefore used here as the leading term in the QCD treatment of the
exclusive decay.

It is important to emphasize that this approximation does not imply that QCD corrections are absent. Beyond leading order, gluon-mediated corrections can modify the short-distance coefficient and may also generate non-factorizable contributions involving the heavy-quark and light-quark sectors. Such effects are not explicitly calculated in the present work. The central values reported below should therefore be
understood as leading-order predictions within the factorization framework.

The use of the factorization approximation is further motivated by the large energy released in the decay of the top quark. For a light final meson, its energy in the top-quark rest frame is of order $m_t/2$,
up to corrections from the final-state masses. Thus, the production of the energetic light-quark pair occurs at a short-distance scale associated with the heavy top-quark decay, whereas the formation of the
meson is described by its nonperturbative decay constant. This separation of short- and long-distance dynamics is the basis of the factorized description adopted in this analysis.

Possible corrections to this leading-order picture, including higher-order QCD and non-factorizable effects, are therefore regarded as theoretical uncertainties rather than being absorbed into the central values of the predictions. Their quantitative impact is discussed separately in the uncertainty analysis.

\section{Numerical results}
\label{sec:results}

In this section we present the numerical predictions for the partial widths and branching ratios of the rare decays $t \to b\, M,$ where $M = P,\, V$ using the factorized amplitudes derived in Sec.\ref{sec:sm}. All results are obtained with the following input parameters \cite{PDG}: 

\begin{align*}
	m_t &= 172.6 \pm 0.27~\text{GeV}, & 	m_b &= 4.186 \pm 0.0006~\text{GeV}, \\
	\Gamma_t^{\rm tot} &= 1.42^{+19} _{-15} ~\text{GeV}, & G_F &= 1.166\times 10^{-5}~\text{GeV}^{-2}
\end{align*}

The decay constants and meson masses for pseudoscalar and vector final mesons are taken from PDG and recent lattice/QCD sum-rule determinations~\cite{PDG, Wang:2015, Chang:2018} and listed in Tables ~\ref{T1} and \ref{T2}.

\begin{table}[ht]
	\renewcommand{\arraystretch}{1.15}
	\begin{center}
		\caption{List of final pseudoscalar mesons ~\cite{PDG, Wang:2015, Chang:2018, Baker:2014}.}
			\label{T1}
		\begin{tabular}{cccc}
			{\footnotesize \textbf{Meson}}&{\footnotesize \textbf{Quark}}&{\footnotesize \textbf{Mass}}&{\footnotesize \textbf{Decay}}\\
			&{\footnotesize \textbf{Content}}&{\footnotesize \textbf{(MeV)}}&{\footnotesize \textbf{Constant (MeV)}} \\
			\hline
			$\pi^+$&$u\bar{d}$& $139.57039 \pm 0.00018 $ &$130.3 \pm 0.3$\\
			\hline
			$K^+$&$u\bar{s}$& $493.677 \pm 0.015$ & $156.1 \pm 0.5$\\
			\hline
			$D^+$&$c\bar{d}$& $1869.66 \pm 0.05 $ & $208 \pm 10$ \\
			\hline
			$B^+$&$u\bar{b}$&$ 5279.41 \pm 0.07$ & $194 \pm 15$ \\
			\hline
			$D^+_s$&$c\bar{s}$& $1968.35 \pm 0.07$ & $240 \pm 10$ \\
			\hline
			$B^+_c$&$c\bar{b}$& $6274.47 \pm 0.32 $ & $528 \pm 0.25$			
		\end{tabular}
	\end{center}
\end{table}

\begin{table}[ht]
	\renewcommand{\arraystretch}{1.15}
	\begin{center}
		\caption{List of final vector mesons ~\cite{PDG, Wang:2015, Chang:2018, Baker:2014}.}
			\label{T2}
		\begin{tabular}{cccc}
			{\footnotesize \textbf{Meson}}&{\footnotesize \textbf{Quark}}&{\footnotesize \textbf{Mass}}&{\footnotesize \textbf{Decay}}\\
			&{\footnotesize \textbf{Content}}&{\footnotesize \textbf{(MeV)}}&{\footnotesize \textbf{Constant (MeV)}}\\
			\hline
			$\rho^+$&$u\bar{d}$&$775.11 \pm 0.34$&$210 \pm 10$\\
			\hline
			$K^{\ast+}$&$u\bar{s}$&$891.88 \pm 0.23$&$204 \pm 7$\\
			\hline
			$D^{\ast+}$&$c\bar{d}$& $2010.27 \pm 0.04$ & $263 \pm 21$\\
			\hline
			$B^{\ast+}$&$u\bar{b}$&$5324.75 \pm 0.20$&$213 \pm 18$\\
			\hline
			$D_s^{\ast+}$&$c\bar{s}$&$ 2112.2 \pm 0.4$ &$308 \pm 21$\\
			\hline
			$B_c^{\ast+}$&$c\bar{b}$&$6274.47 \pm 0.32$&$528 \pm 0.25$				
		\end{tabular}
	\end{center}
\end{table}

The CKM factors entering each channel are:

\[
|V_{tb}V_{ud}| = 0.974,\qquad
|V_{tb}V_{cs}| = 0.973,\qquad
|V_{tb}V_{us}| = 0.224,\qquad
|V_{tb}V_{cd}| = 0.221,
\]

Channels involving $V_{ub}$ and $V_{cb}$ are strongly CKM suppressed and therefore neglected.

The uncertainties associated with the input parameters, including the quark and meson masses, meson decay constants, and CKM matrix elements, are propagated to the calculated partial widths and branching fractions.
The resulting uncertainties are incorporated directly into the results presented in Tables~\ref{T3} and \ref{T4}, where the predictions are given in terms of their central values together with the corresponding
upper and lower limits.

\subsection{Pseudoscalar final states}

The partial width for $t\to bP$ is given in Eq.~\ref{gammap}. Using the numerical inputs, we obtain the results in Table~\ref{T3}.

\begin{table}[h!]
	\centering
	\caption{Predicted partial widths and branching ratios for
		$t\to bP$. The relative deviations are calculated with respect
		to the results reported in Ref.~\cite{dEnterria:2025}.}
	\label{T3}
	\begin{tabular}{cccc}
		Meson & $\Gamma(t\to bP)$ (GeV) & BR$(t\to bP)$ & Relative error\\
		\hline
		$\pi^+$   & $1.13^{\pm0.005} \times 10^{-7}$ & $1.605^{\pm0.007} \times 10^{-7}$ & $0.3\%$ \\
		$K^+$     & $8.552^{\pm0.04} \times 10^{-9}$ & $1.214^{\pm0.006} \times 10^{-8}$ & $1.1\%$ \\
		$D_s^{+}$ & $3.784^{\pm0.018}\times 10^{-7}$ & $5.374^{\pm0.025} \times 10^{-7}$ & $10.3\%$ \\
		$D^{+}$   & $1.52^{\pm0.007} \times 10^{-8}$ & $2.158^{\pm0.011} \times 10^{-8}$ & $6.1\%$ \\
	\end{tabular}
\end{table}

The obtained results are compared with the recent comprehensive study of rare and exclusive few-body decays of the Standard Model particles by d'Enterria and Lê~\cite{dEnterria:2025}. In particular, this study provides updated predictions for semiexclusive top-quark decays into a bottom quark and a light pseudoscalar or vector meson. For the pseudoscalar channels considered here, our results are found to be in good agreement with the updated predictions reported in Ref.~\cite{dEnterria:2025}.

It is worth noting that the predictions for some of these decay modes have been revised in the recent literature. In particular, the branching fraction for $t\to b\pi^+$ reported in earlier work is substantially smaller than the updated result quoted in Ref.~\cite{dEnterria:2025}. The latter study revisits the normalization and factorization treatment of these exclusive top-quark decays and obtains a branching fraction of approximately $1.6\times10^{-7}$ for $t\to b\pi^+$. Our result, $\mathcal{B}(t\to b\pi^+)\simeq1.605\times10^{-7}$, is therefore consistent with this updated prediction.

A similar agreement is observed for the other pseudoscalar channels. The predicted branching fractions for $t\to bK^+$, $t\to bD^+$, and $t\to bD_s^+$ are approximately $1.214\times10^{-8}$, $2.158\times10^{-8}$, and $5.374\times10^{-7}$, respectively, which are close to the corresponding values reported in Ref.~\cite{dEnterria:2025}. This agreement provides an important consistency check of the decay amplitudes, spin averaging, two-body phase-space normalization, and meson decay-constant conventions used in our calculation.

The observed hierarchy $\text{BR}(t\to bD_s) \gg \text{BR}(t\to b\pi^+) \gg \text{BR}(t\to bK^+)$ is primarily determined by the corresponding CKM matrix elements and is consistent with the expectations of previous studies.

\subsection{Vector final states}

For vector mesons, the amplitude contains an additional factor of $m_V$ and a polarization sum.  
The partial width is given in Eq.~\ref{gammav}. Therefore The numerical results are shown in Table~\ref{T4}.

\begin{table}[h!]
	\centering
	\caption{Predicted partial widths and branching ratios for
		$t\to bV$. The relative deviations are calculated with respect
		to the results reported in Ref.~\cite{dEnterria:2025}.}
	\label{T4}
	\begin{tabular}{cccc}
		Meson & $\Gamma(t\to bV)$ (GeV) &
		$\mathrm{BR}(t\to bV)$ & Relative error \\
		\hline
		$\rho^+$      & $2.90405^{\pm0.013} \times10^{-7}$ & $4.12375^{\pm0.019} \times10^{-7}$ & $4.1\%$ \\
		$K^{\ast+}$   & $1.46242^{\pm0.007} \times10^{-8}$ & $2.076664^{\pm0.009} \times10^{-8}$ & $1.4\%$ \\
		$D_s^{\ast+}$ & $6.23412^{\pm0.03} \times10^{-7}$ & $8.85245^{\pm0.04} \times10^{-7}$ & $10.2\%$ \\
		$D^{\ast+}$   & $2.43066^{\pm0.011} \times10^{-8}$ & $3.45154^{\pm0.016} \times10^{-8}$ & $6.7\%$ \\
	\end{tabular}
\end{table}

The predicted partial widths and branching fractions for the vector-meson channels are summarized in Table~\ref{T4}. The results are compared with the updated theoretical predictions presented by d'Enterria and Lê~\cite{dEnterria:2025}, who provide a comprehensive analysis of rare and exclusive few-body decays of Standard Model particles, including exclusive top-quark decays into a bottom quark and a vector meson.

For the $\rho^+$ channel, we obtain $\mathrm{BR}(t\to b\rho^+) = 4.123\times10^{-7}$ which is consistent with the predictions the value of approximately $4.3\times10^{-7}$ reported in Ref.~\cite{dEnterria:2025}, corresponding to a relative deviation of about $4.1\%$. For the $K^{\ast+}$ channel, our prediction, $
\mathrm{BR}(t\to bK^{\ast+}) = 2.07\times10^{-8} $ differs from the corresponding prediction of approximately $2.1\times10^{-8}$ in Ref.~\cite{dEnterria:2025} by about $1.4\%$, which indicates reasonable agreement given the sensitivity of this channel to the input parameters.

For the heavier vector-meson channels, the deviations are more pronounced. We find $\mathrm{BR}(t\to bD^{\ast+}) = 3.45\times10^{-8}$ and $ mathrm{BR}(t\to bD_s^{\ast+}) = 8.85\times10^{-7}$ to be compared with $3.7\times10^{-8}$ and $9.8\times10^{-7}$, respectively, reported in Ref.~\cite{dEnterria:2025}. The corresponding relative deviations are approximately $6.7\%$ and $10.2\%$. These larger differences may arise from the sensitivity of the heavier channels to the meson decay constants, finite-mass effects, and the specific input parameters and approximations adopted in the two calculations.

Overall, the comparison shows very good agreement for the $\rho^+$ channel and reasonable agreement for the $K^{\ast+}$ channel, while the heavier $D^{\ast+}$ and $D_s^{\ast+}$ modes exhibit larger deviations. Nevertheless, the comparison provides a useful consistency check of the normalization of the $V-A$ decay amplitude, the initial-state spin averaging, the vector-meson polarization sum, and the two-body phase-space factor used in the present calculation. In particular, the comparison with the updated predictions supports the correction of the factor-of-four normalization discrepancy in the original numerical results, indicating that the discrepancy originated from the decay-width normalization rather than from the Dirac-trace calculation itself.

It should be noted that the comparison with the results of ref.\cite{dEnterria:2025} should be interpreted with some caution, since the theoretical frameworks and the treatment of higher-order QCD effects are not necessarily identical. The results presented in this work are obtained within the leading-order factorization approximation, whereas the reference analysis may incorporate additional theoretical ingredients and uncertainties. Therefore, the comparison is intended primarily to assess the consistency of the predicted order of magnitude and the overall phenomenological behavior rather than to imply numerical equivalence
between the two approaches.

\section{Helicity amplitudes and polarization fractions for vector final states}
\label{sec:vector-helicity}

For vector final states, $t\to bV$, the polarization of the meson $V$ provides additional observables beyond the total rate. To access polarization observables we decompose the rate into helicity components, and define the helicity fraction $F_\lambda$ 

\begin{equation}
	\Gamma(t\to bV)=\Gamma_0+\Gamma_-+\Gamma_+, \qquad
	F_\lambda \equiv \frac{\Gamma_\lambda}{\Gamma_0+\Gamma_-+\Gamma_+},
	\qquad \lambda=0,\pm
	\label{gammavh}
\end{equation}

We work in the top-quark rest frame and choose the $z$ axis along the $V$ three-momentum. So we have

\begin{equation}
	p_t^\mu=(m_t,0,0,0),\qquad
	p_V^\mu=(E_V,0,0,+|\vec p\,|),\qquad
	p_b^\mu=(E_b,0,0,-|\vec p\,|)
	\label{mumenum3}
\end{equation}

with

\begin{equation}
	\begin{aligned}
		E_V &= \frac{m_t^2+m_V^2-m_b^2}{2m_t}, &
		E_b &= \frac{m_t^2+m_b^2-m_V^2}{2m_t}, \\
		|\vec p\,| &= \frac{1}{2m_t}\,\lambda^{1/2}(m_t^2,m_b^2,m_V^2), &
		\lambda(x,y,z) &= x^2+y^2+z^2-2xy-2xz-2yz
	\end{aligned}
	\label{energy}
\end{equation}

For a massive particle moving along $+z$ we use the helicity basis ($g^{\mu\nu}=\mathrm{diag}(1,-1,-1,-1)$)

\begin{equation}
	\begin{aligned}
		\varepsilon^\mu(0) &= \left(\frac{|\vec p\,|}{m_V},\,0,\,0,\,\frac{E_V}{m_V}\right),\\
		\varepsilon^\mu(\pm) &= \frac{1}{\sqrt{2}}(0,\,\mp 1,\,-i,\,0)
	\end{aligned}
	\label{epsilon}
\end{equation}

The longitudinal vector in Eq.~\ref{epsilon} follows directly from the ansatz $\varepsilon^\mu(0)=(a,0,0,b)$ combined with the constraints $p_V\!\cdot\!\varepsilon(0)=0$ and $\varepsilon(0)^2=-1$, which uniquely fix $a=|\vec p\,|/m_V$ and $b=E_V/m_V$.

The helicity amplitudes are defined by

\begin{equation}
	\mathcal{A}_\lambda =\frac{G_F}{\sqrt{2}}\,V_{tb}V_{qq'}^\ast\,	f_V m_V\,
	\big[ \bar u_b(p_b)\gamma^\mu(1-\gamma_5)u_t(p_t)\; \varepsilon_\mu^\ast(\lambda) \big]
	\label{HelAmp}
\end{equation}

and the corresponding helicity widths are

\begin{equation}
	\Gamma_\lambda(t\to bV_\lambda) =\frac{1}{16\pi m_t^3}\, \lambda^{1/2}(m_t^2,m_b^2,m_V^2)\;
	\overline{|\mathcal{A}_\lambda|^2}
	\label{Helwidth}
\end{equation}

where the bar denotes the spin sum over the final $b$ quark and the average over the initial top-quark spin, consistent with the normalization adopted in Sec.~\ref{sec:sm}.

In the phenomenologically relevant limit $m_b\to 0$, the SM charged-current vertex is purely left-chiral,
$\gamma^\mu(1-\gamma_5)=2\gamma^\mu P_L$. Therefore the produced $b$ quark is left-chiral. In the limit $m_b\to 0$, chirality coincides with helicity and the $b$ is produced with negative helicity

\begin{equation}
	m_b\to 0
	\qquad \Rightarrow \qquad
	b_L \equiv P_L b \;\; \Longleftrightarrow \;\; h_b=-\frac12
	\label{mb0}
\end{equation}

The helicity structure of the decay follows from the left-chiral charged-current interaction. The decay amplitude for a vector meson with helicity $\lambda$ is obtained by projecting the weak current onto the corresponding polarization state,

\begin{equation}
	\begin{alignedat}{2}
		\mathcal A_\lambda &= C_V\,J^\mu\,\varepsilon_\mu^\ast(\lambda)
		\qquad\qquad
		&J^\mu &= \bar u_b(p_b)\gamma^\mu(1-\gamma_5)u_t(p_t) \\
		&&C_V &= \frac{G_F}{\sqrt{2}}\, V_{tb}V_{qq'}^\ast\,f_V m_V
	\end{alignedat}
	\label{AJC}
\end{equation}

In the top-quark rest frame we choose the $z$ axis along the vector-meson momentum, $\vec p_V\parallel +\hat z$, so that the $b$ quark moves in the opposite direction, $\vec p_b\parallel -\hat z$. In the limit $m_b\to0$, the weak interaction produces a left-chiral $b$ quark, which is equivalent to a negative-helicity state. Since the spin of a negative-helicity massless fermion is opposite to its momentum, the $b$ quark carries

\begin{equation}
	\vec p_b \parallel -\hat z, \qquad h_b=-\frac12 \qquad\Rightarrow\qquad s_{b,z}=+\frac12
	\label{phs}
\end{equation}

Angular-momentum conservation along the $z$ axis then determines the allowed vector-meson helicities. The configuration with $\lambda=+1$ would require a total final-state angular momentum projection $J_z=+3/2$, which cannot be obtained from the decay of a spin-$1/2$ top quark. Consequently, the right-handed transverse helicity amplitude vanishes in the massless-$b$ limit,

\begin{equation}
	\mathcal A_+^{\rm SM}(m_b=0)=0 	\qquad\Longrightarrow\qquad \Gamma_+^{\rm SM}(m_b=0)=0
	\label{a-gamma}
\end{equation}

up to corrections of order ${\cal O}(m_b^2/m_t^2)$ for finite $m_b$.

\subsection{Helicity amplitude for Finite \texorpdfstring{$\boldsymbol{m_b}$}{mb}}
\label{sec:vector-helicity-mb}

In Sec.~\ref{sec:vector-helicity}, we have shown that in the limit $m_b \to 0$, where chirality and helicity coincide, the SM amplitude satisfies $\mathcal{A}_+^{\rm SM}(m_b=0)=0$. In this section, we extend the analysis to the case of a finite bottom-quark mass. In this more general framework, the exact correspondence between chirality and helicity no longer holds, and the helicity-suppressed $\lambda=+1$ amplitude can receive a non-vanishing contribution. We compute the helicity amplitudes starting from the exact spin-summed tensor and systematically quantify the effects arising from finite-$m_b$ corrections, verifying the persistence and structure of the suppression beyond the massless limit.

we work in the top-quark rest frame and choose the $z$ axis along the vector-meson momentum, and the helicity amplitudes are defined in Eq.~\ref{AJC}. After summing over final-state spins and averaging over the initial top-quark spin, the squared amplitudes can be written as

\begin{equation}
	\overline{|\mathcal A_\lambda|^2} =	C_V^2\,L_{\mu\nu}\, \varepsilon^{\ast\mu}(\lambda)\,
	\varepsilon^\nu(\lambda)
	\label{a-lambda}
\end{equation}

where $L_{\mu\nu}$ is the hadronic tensor and defined by Eq.~\ref{LmunuV}. Evaluating the Dirac trace yields

\begin{equation}
	L_{\mu\nu} =4\Big[ p_{b\mu}p_{t\nu} +p_{b\nu}p_{t\mu}-g_{\mu\nu}(p_b\!\cdot p_t)	+i\epsilon_{\mu\nu\alpha\beta}p_b^\alpha p_t^\beta \Big]
	\label{Lmunu_correct}
\end{equation}

For the transverse polarization vectors

\begin{equation}
	\varepsilon^\mu(\pm)=\frac{1}{\sqrt{2}}(0,\mp1,-i,0)
\end{equation}

the contractions give

\begin{equation}
	L_\pm \equiv L_{\mu\nu}\varepsilon^{\ast\mu}(\pm)\varepsilon^\nu(\pm) =4m_t(E_b \mp p)
\end{equation}

so that

\begin{equation}
	\overline{|\mathcal A_\pm|^2} =	\mathcal C_V^2\,4m_t(E_b \mp p)
\end{equation}

Using the kinematic identity

\begin{equation}
	E_b^2-p^2=m_b^2, 	\qquad 	E_b-p=\frac{m_b^2}{E_b+p}
\end{equation}

one obtains

\begin{equation}
	\overline{|\mathcal A_+|^2} =\mathcal C_V^2\,4m_t\,\frac{m_b^2}{E_b+p}
\end{equation}

which makes the helicity suppression explicit $\overline{|\mathcal A_+|^2}\propto m_b^2$. Thus,

\begin{equation}
	\overline{|\mathcal A_+|^2}\to0 \qquad (m_b\to0)
\end{equation}

For the longitudinal polarization vector

\begin{equation}
	\varepsilon^\mu(0)=\left(\frac{p}{m_V},0,0,\frac{E_V}{m_V}\right)
\end{equation}

one finds

\begin{equation}
	L_0 =\frac{2}{m_V^2} \Big[(m_t^2-m_b^2)^2 - m_V^2(m_t^2+m_b^2) \Big]
\end{equation}

The longitudinal amplitude is then

\begin{equation}
	\overline{|\mathcal A_0|^2} = \mathcal C_V^2\,L_0
\end{equation}

The detailed derivation of the pseudoscalar Dirac trace is presented in Appendix~\ref{app:helicity}.

\subsection{Helicity fractions at finite \texorpdfstring{$\boldsymbol{m_b}$}{mb}}

Since all helicity channels share the same two body phase space factor, the helicity fractions can be written directly in terms of the squared amplitudes:

\begin{equation}
	F_\lambda = \frac{|\mathcal A_\lambda|^2} {|\mathcal A_0|^2+|\mathcal A_-|^2+|\mathcal A_+|^2}
\end{equation}

Defining

\begin{equation}
	N \equiv (m_t^2-m_b^2)^2 - m_V^2(m_t^2+m_b^2)
\end{equation}

the exact closed-form expressions are

\begin{equation}
	F_0 = \frac{N/m_V^2} {N/m_V^2 + 4m_t E_b}
\end{equation}

\begin{equation}
	F_- =\frac{2m_t(E_b+p)}	{N/m_V^2 + 4m_t E_b}
\end{equation}

\begin{equation}
	F_+	=\frac{2m_t(E_b-p)}	{N/m_V^2 + 4m_t E_b}=\frac{2m_t}{N/m_V^2 + 4m_t E_b}\,\frac{m_b^2}{E_b+p}
\end{equation}

These expressions make the helicity suppression of the right-handed mode explicit:

\begin{equation}
	F_+ \propto {m_b^2}
\end{equation}

In the limit $m_b\to0$ one recovers

\begin{equation}
	F_+\to0,\qquad
	F_-\to \frac{2m_V^2}{m_t^2+2m_V^2},\qquad
	F_0\to \frac{m_t^2}{m_t^2+2m_V^2}
\end{equation}

and therefore

\begin{equation}
	\frac{\Gamma_0}{\Gamma_-}\simeq \frac{m_t^2}{2m_V^2}
\end{equation}

Table~\ref{helicity_fractions_finite_mb} summarizes the SM helicity fractions for $t\to bV$ decays including finite bottom-quark mass effects. The numerical evaluation is performed using $m_t=172.6 \pm 0.27~\mathrm{GeV}$ and $m_b=4.18\pm 0.0006~\mathrm{GeV}$, together with the physical meson masses.

The uncertainties of the relevant input parameters were propagated to the helicity fractions to quantify their numerical stability. The corresponding upper and lower bounds indicate that the predicted values vary only mildly within the allowed parameter ranges. In particular, the suppression of the right-handed fraction $F_+$ persists across the considered input variations. No appreciable variation was observed, in particular for the suppressed right-handed fraction $F_+$, indicating that the small helicity fractions are stable against the quoted input uncertainties.

\begin{table}[h!]
	\centering
	\caption{Finite-$m_b$ SM helicity fractions for $t\to bV$ decays.}
	\label{helicity_fractions_finite_mb}
	\begin{tabular}{lccc}
		$V$ & $F_0$ & $F_-$ & $F_+$ \\
		\hline
		$\rho^+$        & 0.99996  & $4.03691^{\pm0.01}\times10^{-5}$ & $2.39047^{\pm0.01}\times10^{-8}$ \\
		$K^{\ast+}$     & 0.999947 & $5.33575^{\pm0.01}\times10^{-5}$ & $3.15962^{\pm0.01} \times10^{-8}$ \\
		$D^{\ast+}$     & 0.999729 & $2.70631^{\pm0.008}\times10^{-4}$   & $1.59792^{\pm0.01}\times10^{-7}$ \\
		$D_s^{\ast+}$   & 0.99973 & $2.69556^{\pm0.008}\times10^{-4}$  & $1.59157^{\pm0.01}\times10^{-7}$
	\end{tabular}
\end{table}

The numerical results in Table~\ref{helicity_fractions_finite_mb} exhibit a clear hierarchy among the three polarization fractions. For all considered vector-meson channels, the longitudinal component dominates, with $F_0\simeq1$, while the left-handed transverse contribution is strongly suppressed, $F_-\ll F_0$. The right-handed fraction is further suppressed by several orders of magnitude, $F_+\ll F_-$, in agreement with the analytical behavior $F_+\propto m_b^2$. This strong suppression originates from the left-handed $V-A$ structure of the charged-current interaction and the mass-suppressed helicity flip required for the right-handed configuration.

\begin{figure}[t]
	\centering
	\includegraphics[width=0.85\linewidth]{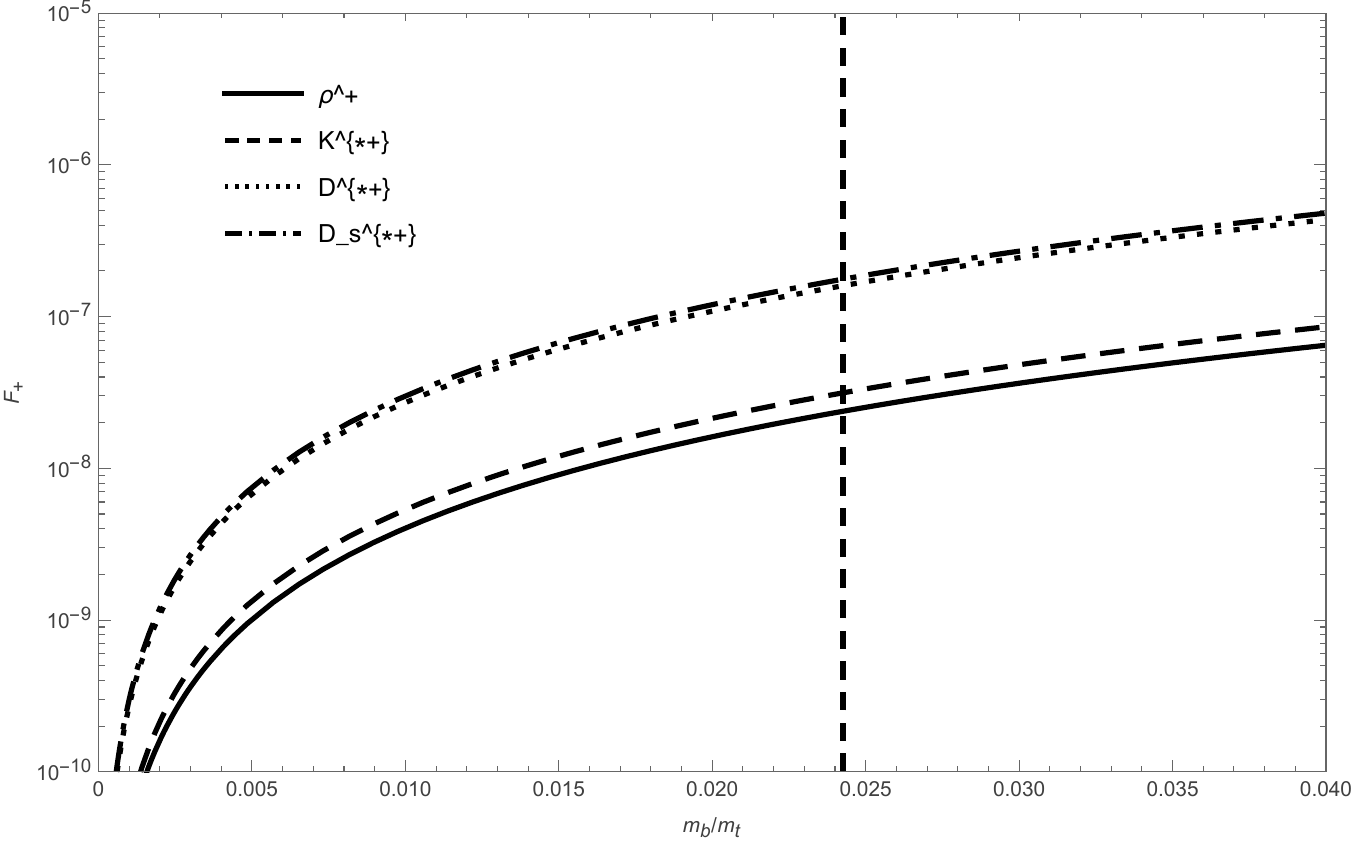}
	\caption{Dependence of the right-handed helicity fraction $F_+$ on the mass ratio $m_b/m_t$ for the $\rho^+$, $K^{*+}$, $D^{*+}$, and $D_s^{*+}$ channels. The physical value $m_b/m_t\simeq0.0243$ corresponds to the Standard Model input parameters used in this analysis.}
	\label{Fplus_mb_mt}
\end{figure}

The dependence of the right-handed helicity fraction $F_+$ on the mass ratio $m_b/m_t$ is shown in Fig.~\ref{Fplus_mb_mt} for the four considered vector-meson channels. In the limit $m_b/m_t \to 0$, all channels exhibit $F_+ \to 0$, reflecting the suppression of the right-handed helicity amplitude in the massless-$b$ limit. For small $m_b/m_t$, the dependence is approximately quadratic, 

\begin{equation}
	F_+ \propto \left(\frac{m_b}{m_t}\right)^2.
\end{equation}

At the physical value $m_b/m_t \simeq 0.0243$, $F_+$ remains of order $10^{-8}$--$10^{-7}$, with the $D^{*+}$ and $D_s^{*+}$ channels yielding slightly larger values than the $\rho^+$ and $K^{*+}$ channels. The
persistence of this strong suppression over the physically relevant range indicates that the small predicted values of $F_+$ are robust against variations in the $b$-quark mass within the considered kinematic region.

The dependence on the vector-meson mass is also consistent with the massless-bottom limit. As $m_V$ increases, the longitudinal fraction decreases while the transverse left-handed fraction increases, as
suggested by
\[
F_0 \simeq \frac{m_t^2}{m_t^2+2m_V^2},
\qquad
F_- \simeq \frac{2m_V^2}{m_t^2+2m_V^2}.
\]
The numerical results therefore reproduce the expected polarization hierarchy and its dependence on the final-state meson mass.

Note that the stability of the helicity fractions was checked by varying $m_t$ and $m_b$ within their quoted uncertainties and repeating the calculation of the polarized decay widths. These variations are already included in the upper and lower bounds of the numerical predictions. No appreciable variation in the resulting helicity fractions was observed. This weak sensitivity is expected from the hierarchy $m_t \gg m_b,m_V$. In particular, the strongly suppressed right-handed fraction $F_+$ exhibits only a small variation under the allowed changes in $m_t$ and $m_b$, indicating that the predicted suppression is not sensitive to these input-parameter uncertainties.

\section{Discussion}
\label{sec:discussion}

The factorized framework adopted in this work is based on the Standard Model charged-current interaction and provides a systematic description of the exclusive decays $t\to bM$. Within this framework, the hierarchy among the different decay channels is primarily governed by the corresponding CKM matrix elements, together with the meson decay constants, phase-space factors, and the relevant kinematic functions. In particular, the channels associated with the largest CKM factors, such as $t\to bD_s^{(\ast)}$, are expected to have the largest branching fractions, whereas transitions involving the suppressed $V_{td}$ and $V_{ts}$ elements are significantly reduced.

The treatment of the off-shell $W$ propagator is justified by the hierarchy $m_M^2\ll m_W^2$ for all mesons considered in this work. Consequently, the propagator can be expanded in powers of $m_M^2/m_W^2$, and the leading term reproduces the effective four-fermion interaction used in the factorized amplitudes. The corresponding corrections are numerically negligible at the accuracy considered here.

For vector-meson final states, the polarization structure provides additional information beyond the total decay rate. The decay amplitude can be decomposed into longitudinal and transverse helicity components, allowing the corresponding helicity fractions to be studied independently of the overall normalization to a large extent. In particular, the helicity observables are less sensitive to common normalization factors such as CKM matrix elements and meson decay constants, making them useful complementary observables to the total branching fractions.

The results obtained in this work therefore provide a consistent Standard Model baseline for exclusive pseudoscalar and vector meson production in top-quark decays. The explicit trace calculations and helicity decomposition presented in the appendices also make the normalization and kinematic dependence of the results transparent and provide a basis for future extensions involving higher-order QCD corrections or non-standard charged-current interactions.

\section{Conclusion}
\label{sec:conclusion}

In this work, a leading-order analysis of rare exclusive decays $t \to bM$, where $M$ denotes a pseudoscalar or vector meson, has been presented. Starting from the underlying partonic transition $t \to bW^\ast \to b(q\bar q')$ and employing the standard definitions of meson decay constants, compact factorized expressions for the decay widths of the $t \to bP$ and $t \to bV$ channels have been derived.

The numerical results indicate branching ratios in the range $\mathcal{O}(10^{-8})$--$\mathcal{O}(10^{-7})$, with a clear hierarchy among the different channels governed primarily by the relevant CKM matrix elements and meson decay constants. The results should be regarded as leading-order predictions, with theoretical uncertainties associated with the non-perturbative hadronic inputs, scale dependence, and higher-order QCD corrections.

For vector final states, the helicity structure has been analyzed in detail. In the Standard Model and in the limit $m_b \to 0$, the polarization fractions take the form
\[
F_0^{\rm SM} \simeq \frac{m_t^2}{m_t^2 + 2m_V^2},
\qquad
F_-^{\rm SM} \simeq \frac{2m_V^2}{m_t^2 + 2m_V^2},
\qquad
F_+^{\rm SM} \simeq 0
\]
Finite bottom-quark mass effects generate a non-vanishing but highly suppressed right-handed contribution, with the leading suppression controlled by $m_b^2/m_t^2$.

Overall, the results provide a consistent Standard Model baseline for rare exclusive top-quark decays within the adopted factorization framework. The analysis can be extended by incorporating higher-order QCD corrections, improved non-perturbative inputs, and a more systematic treatment of factorization effects. The helicity observables also provide a useful starting point for future studies of non-standard $tWb$ interactions within an effective field theory framework.

\appendix

\section{Evaluation of the Dirac Trace for Pseudoscalar Meson Final States}
\label{app:pseudoscalar}

The decay amplitude for the pseudoscalar meson production is given by

\begin{equation}
	\mathcal{A} =C \, p_P^\mu \bar u_b(p_b) \gamma_\mu(1-\gamma_5) u_t(p_t)
\end{equation}

where

\begin{equation}
	C= \frac{G_F}{\sqrt2} V_{tb} V_{qq'}^\ast f_P
\end{equation}

After summing over the spins of the final-state particles and averaging over the initial top-quark spin, the squared amplitude becomes

\begin{equation}
	\overline{\sum} |\mathcal A|^2 = |C|^2 p_P^\mu p_P^\nu L_{\mu\nu}
\end{equation}

where the Dirac tensor is defined as

\begin{equation}
	L_{\mu\nu} = \frac12 \mathrm{Tr} \left[ (\slashed p_b+m_b) \gamma_\mu (1-\gamma_5) (\slashed p_t+m_t) \gamma_\nu (1-\gamma_5) \right]
\end{equation}

The factor $1/2$ originates from averaging over the two spin states of the initial top quark.

The left-handed projection operators are now expanded explicitly,

\begin{align}
	L_{\mu\nu} =& \frac12 \mathrm{Tr} \Big[ (\slashed p_b+m_b) \gamma_\mu (\slashed p_t+m_t) \gamma_\nu \Big] 
	\nonumber\\ & -\frac12 \mathrm{Tr} \Big[ (\slashed p_b+m_b) \gamma_\mu \gamma_5 (\slashed p_t+m_t) \gamma_\nu \Big]
	\nonumber\\ & -\frac12 \mathrm{Tr} \Big[ (\slashed p_b+m_b) \gamma_\mu (\slashed p_t+m_t) \gamma_\nu \gamma_5 \Big] 
	\nonumber\\ & +\frac12 \mathrm{Tr} \Big[ (\slashed p_b+m_b) \gamma_\mu \gamma_5 (\slashed p_t+m_t) \gamma_\nu \gamma_5 \Big]
\end{align}

For convenience we define

\begin{align}
	T_1 &= \mathrm{Tr} \Big[ (\slashed p_b+m_b) \gamma_\mu (\slashed p_t+m_t) \gamma_\nu \Big] 	\\
	T_2 &= \mathrm{Tr} \Big[ (\slashed p_b+m_b) \gamma_\mu \gamma_5 (\slashed p_t+m_t) \gamma_\nu \Big] \\
	T_3 &= \mathrm{Tr} \Big[ (\slashed p_b+m_b) \gamma_\mu (\slashed p_t+m_t) \gamma_\nu \gamma_5 \Big] \\
	T_4 &= \mathrm{Tr} \Big[ (\slashed p_b+m_b) \gamma_\mu \gamma_5 (\slashed p_t+m_t) \gamma_\nu \gamma_5 \Big]
\end{align}

such that

\begin{equation}
	L_{\mu\nu} = \frac12 \left( T_1-T_2-T_3+T_4 \right)
\end{equation}

\subsection*{Evaluation of $T_1$}

We first evaluate the vector contribution $T_1$. Expanding both fermion mass terms gives

\begin{align}
	T_1 = \mathrm{Tr} \left[ \slashed p_b \gamma_\mu \slashed p_t \gamma_\nu \right] +m_t \mathrm{Tr} \left[ \slashed p_b \gamma_\mu \gamma_\nu \right] +m_b \mathrm{Tr} \left[ \gamma_\mu \slashed p_t \gamma_\nu \right] +m_tm_b \mathrm{Tr} \left[  \gamma_\mu \gamma_\nu \right]
\end{align}

The second and third terms vanish identically because the trace of an odd number of Dirac gamma matrices is zero,

\begin{equation}
	\mathrm{Tr} (\gamma^\alpha \gamma^\beta \gamma^\gamma) = 0
\end{equation}

\begin{equation}
	\mathrm{Tr} ( \gamma^\alpha \gamma^\mu \gamma^\beta \gamma^\nu ) = 4 \left( g^{\alpha\mu} g^{\beta\nu} - g^{\alpha\beta} g^{\mu\nu} + g^{\alpha\nu} g^{\mu\beta} \right)
\end{equation}

Therefore,

\begin{equation}
	T_1 = \mathrm{Tr} [ \slashed p_b \gamma_\mu \slashed p_t \gamma_\nu ] + m_tm_b \mathrm{Tr} ( \gamma_\mu
	\gamma_\nu )
\end{equation}

Finally,

\begin{equation}
	T_1 = 4 \left( p_{b\mu} p_{t\nu} + p_{b\nu} p_{t\mu} - g_{\mu\nu} p_b\!\cdot p_t \right) + 4m_tm_b g_{\mu\nu}
\end{equation}

\subsection*{Evaluation of $T_2$}

We next evaluate the axial-vector contribution

\begin{equation}
	T_2= \mathrm{Tr} \left[ (\slashed p_b+m_b) \gamma_\mu\gamma_5 (\slashed p_t+m_t) \gamma_\nu \right]
\end{equation}

Expanding the mass terms,

\begin{equation}
	T_2 = \mathrm{Tr} \left[ \slashed p_b \gamma_\mu\gamma_5 \slashed p_t \gamma_\nu \right]
	+m_t \mathrm{Tr} \left[ \slashed p_b \gamma_\mu\gamma_5 \gamma_\nu \right]
	+m_b \mathrm{Tr} \left[ \gamma_\mu\gamma_5 \slashed p_t \gamma_\nu \right]
	+m_tm_b \mathrm{Tr} \left[ \gamma_\mu\gamma_5 \gamma_\nu \right]
\end{equation}

The last three traces vanish because they contain fewer than four Dirac gamma matrices together with one $\gamma_5$. Therefore,

\begin{equation}
	T_2 = \mathrm{Tr} \left[ \slashed p_b \gamma_\mu\gamma_5 \slashed p_t \gamma_\nu \right].
\end{equation}

Using

\begin{equation}
	\mathrm{Tr} ( \gamma^\alpha \gamma^\mu \gamma^\beta \gamma^\nu \gamma_5 ) = -4i \epsilon^{\alpha\mu\beta\nu}
\end{equation}

one finds

\begin{equation}
	T_2 = -4i \epsilon_{\mu\nu\alpha\beta} p_b^\alpha p_t^\beta
\end{equation}

\subsection*{Evaluation of $T_3$}

The third trace is

\begin{equation}
	T_3= \mathrm{Tr} \left[ (\slashed p_b+m_b) \gamma_\mu (\slashed p_t+m_t) \gamma_\nu \gamma_5 \right]
\end{equation}

Expanding the mass terms,

\begin{equation}
	T_3 = \mathrm{Tr} \left[ \slashed p_b \gamma_\mu \slashed p_t \gamma_\nu 	\gamma_5 \right]
	+m_t \mathrm{Tr} \left[ \slashed p_b \gamma_\mu \gamma_\nu \gamma_5 \right] 
	+m_b \mathrm{Tr} \left[	\gamma_\mu \slashed p_t \gamma_\nu \gamma_5 \right]
	+m_tm_b \mathrm{Tr} \left[ \gamma_\mu \gamma_\nu \gamma_5 \right]
\end{equation}

Again, the last three traces vanish identically. Therefore,

\begin{equation}
	T_3 = -4i \epsilon_{\mu\nu\alpha\beta} p_b^\alpha p_t^\beta
\end{equation}

\subsection*{Evaluation of $T_4$}

Finally, we evaluate the forth trace,

\begin{equation}
	T_4= \mathrm{Tr} \left[ (\slashed p_b+m_b) \gamma_\mu \gamma_5 (\slashed p_t+m_t) \gamma_\nu \gamma_5 \right]
\end{equation}

Expanding all terms,

\begin{align}
	T_4 =& \mathrm{Tr} \left[ \slashed p_b \gamma_\mu \gamma_5 \slashed p_t \gamma_\nu \gamma_5 \right]
	\nonumber\\ & +m_t \mathrm{Tr} \left[ \slashed p_b \gamma_\mu \gamma_5 \gamma_\nu \gamma_5 \right]
	\nonumber\\ & +m_b \mathrm{Tr} \left[ \gamma_\mu \gamma_5 \slashed p_t \gamma_\nu \gamma_5 \right]
	\nonumber\\ & +m_t m_b \mathrm{Tr} \left[ \gamma_\mu \gamma_5 \gamma_\nu \gamma_5 \right]
\end{align}

The second and third traces vanish because they contain an odd number of Dirac gamma matrices. Using

\begin{equation}
	\{\gamma^\mu,\gamma_5\}=0   \Longrightarrow   \gamma_5\gamma^\mu=-\gamma^\mu \gamma_5
	\label{gamma5_anti}
\end{equation}

the $T_4$ terms become

\begin{equation}
	\mathrm{Tr} ( \slashed p_b \gamma_\mu \gamma_5 \slashed p_t \gamma_\nu \gamma_5 ) = \mathrm{Tr} ( \slashed p_b \gamma_\mu \slashed p_t \gamma_\nu )
\end{equation}

\begin{equation}
	m_t \mathrm{Tr} \left[ \slashed p_b \gamma_\mu \gamma_5 \gamma_\nu \gamma_5 \right]=
	-m_t \mathrm{Tr} \left[ \slashed p_b \gamma_\mu \gamma_\nu \right]=0
\end{equation}

\begin{equation}
	m_b \mathrm{Tr} \left[ \gamma_\mu \gamma_5 \slashed p_t \gamma_\nu \gamma_5 \right]=
	+m_b \mathrm{Tr} \left[ \gamma_\mu \slashed p_t \gamma_\nu \right]=0
\end{equation}

\begin{equation}
	m_t m_b \mathrm{Tr} ( \gamma_\mu \gamma_5 \gamma_\nu \gamma_5 ) = -m_t m_b \mathrm{Tr} (\gamma_\mu \gamma_\nu) = -4g_{\mu\nu}m_t m_b
\end{equation}

Therefore,

\begin{equation}
	T_4 = 4 \left( p_{b\mu}p_{t\nu} +p_{b\nu}p_{t\mu} -g_{\mu\nu}p_b\!\cdot p_t \right) -4m_tm_b g_{\mu\nu}
\end{equation}

Having evaluated all four traces, the Dirac tensor is obtained from

\begin{equation}
	L_{\mu\nu} = \frac12 \left( T_1-T_2-T_3+T_4 \right)
\end{equation}

Substituting the four contributions gives

\begin{align}
	L_{\mu\nu} =& \frac12 \Bigg[ 4\left( p_{b\mu}p_{t\nu} +p_{b\nu}p_{t\mu} -g_{\mu\nu}p_b\!\cdot p_t \right)
	+4m_tm_b g_{\mu\nu} 
	\nonumber\\ & \qquad -\left( -4i \epsilon_{\mu\nu\alpha\beta} p_b^\alpha p_t^\beta \right) -\left( -4i \epsilon_{\mu\nu\alpha\beta} p_b^\alpha p_t^\beta \right)
	\nonumber\\ & \qquad +4\left( p_{b\mu}p_{t\nu} +p_{b\nu}p_{t\mu} -g_{\mu\nu}p_b\!\cdot p_t \right) -4m_tm_b g_{\mu\nu} \Bigg]
\end{align}

Therefore, the final expression for the Dirac tensor is

\begin{equation}
	L_{\mu\nu} 	=  4\left( p_{b\mu}p_{t\nu} +p_{b\nu}p_{t\mu} - g_{\mu\nu} p_b\!\cdot p_t \right) + 4i \epsilon_{\mu\nu\alpha\beta} p_b^\alpha p_t^\beta
\end{equation}

The spin-averaged squared amplitude is

\begin{equation}
	\overline{\sum} |\mathcal A|^2 = |C|^2 p_P^\mu p_P^\nu L_{\mu\nu}
\end{equation}

Substituting this expression into the squared amplitude gives

\begin{align}
	p_P^\mu p_P^\nu L_{\mu\nu} =& 4 p_P^\mu p_P^\nu \left( p_{b\mu}p_{t\nu} + p_{b\nu}p_{t\mu} - g_{\mu\nu} p_b\!\cdot p_t \right) 
	\nonumber\\ & + 4i p_P^\mu p_P^\nu \epsilon_{\mu\nu\alpha\beta} p_b^\alpha p_t^\beta
\end{align}

Since the tensor $ p_P^\mu p_P^\nu $ is symmetric under the interchange $\mu\leftrightarrow\nu$, whereas $\epsilon_{\mu\nu\alpha\beta}$ is antisymmetric, the second term vanishes identically,

\begin{equation}
	p_P^\mu p_P^\nu \epsilon_{\mu\nu\alpha\beta} = 0
\end{equation}

Therefore, only the symmetric part contributes,

\begin{align}
	p_P^\mu p_P^\nu L_{\mu\nu} =& 4 (p_P\!\cdot p_b) (p_P\!\cdot p_t) + 4 (p_P\!\cdot p_b) (p_P\!\cdot p_t) \nonumber\\ & - 4 p_P^\mu p_P^\nu g_{\mu\nu} (p_b\!\cdot p_t)
\end{align}

Using

\begin{equation}
	({p^\mu p_{\mu}})_{P} = p_P^2 = m_P^2
\end{equation}

one obtains

\begin{equation}
	p_P^\mu p_P^\nu L_{\mu\nu} = 8 (p_P\!\cdot p_b) (p_P\!\cdot p_t) - 4 m_P^2 (p_b\!\cdot p_t)
\end{equation}

Hence, the spin-averaged squared amplitude is

\begin{equation}
	\overline{\sum} |\mathcal A|^2 	= 4|C|^2 \left[ 2 (p_P\!\cdot p_b) (p_P\!\cdot p_t) - m_P^2 (p_b\!\cdot p_t) \right]
\end{equation}

Using energy-momentum conservation,

\begin{equation}
	p_t=p_b+p_P
\end{equation}

together with the on-shell conditions

\[
p_t^2=m_t^2,\qquad
p_b^2=m_b^2,\qquad
p_P^2=m_P^2
\]

the required scalar products become

\begin{align}
	p_b\!\cdot p_t &= \frac{m_t^2+m_b^2-m_P^2}{2} \\
	p_P\!\cdot p_t &= \frac{m_t^2-m_b^2+m_P^2}{2} \\
	p_P\!\cdot p_b &= \frac{m_t^2-m_b^2-m_P^2}{2}
	\label{kinematicrel}
\end{align}

Substituting these relations into the squared amplitude yields

\begin{align}
	\overline{\sum} |\mathcal A|^2 &= 4|C|^2 \left[ \frac{ (m_t^2-m_b^2-m_P^2) (m_t^2-m_b^2+m_P^2) }{2} - \frac{ 	m_P^2 (m_t^2+m_b^2-m_P^2) }{2} \right] 
	\nonumber\\ &= 2|C|^2 \left[ (m_t^2-m_b^2)^2 - m_P^2 (m_t^2+m_b^2) \right]
\end{align}

Therefore, 

\begin{equation}
	\mathcal{K}_P(m_t,m_b,m_P)= (m_t^2-m_b^2)^2 - m_P^2 (m_t^2+m_b^2) 
\end{equation}

\section{Evaluation of the Dirac Traces for Vector Meson Final States}
\label{app:vector}

For vector meson production, the decay amplitude is

\begin{equation}
	\mathcal{A}(t\to bV_\lambda) = \frac{G_F}{\sqrt2} V_{tb} V_{qq'}^\ast  f_V m_V \, \varepsilon^\ast_\mu (P_V, \lambda) \, \bar u_b(p_b) \gamma^\mu(1-\gamma_5) u_t(p_t)
	\label{eq:AvecApp}
\end{equation}

After summing over the vector-meson polarizations \textbf{}and averaging over the initial top-quark spin,

\begin{equation}
	\overline{\sum} |\mathcal A|^2 = |C|^2 P_{\mu\nu} L^{\mu\nu}
	\label{eq:AvecSqApp}
\end{equation}

where

\begin{equation}
	C= \frac{G_F}{\sqrt2}V_{tb}V_{qq'}^\ast f_V m_V
\end{equation}

and

\begin{equation}
	P_{\mu\nu} =\varepsilon^\ast_\mu (P_V, \lambda) \varepsilon_\nu (P_V, \lambda) = -g_{\mu\nu} + \frac{p_{V\mu}p_{V\nu}}{m_V^2}
	\label{PolSumApp}
\end{equation}

The leptonic tensor is

\begin{equation}
	L_{\mu\nu} = \frac12 \mathrm{Tr} \left[ (\slashed p_b+m_b) \gamma_\mu(1-\gamma_5) (\slashed p_t+m_t) \gamma_\nu(1-\gamma_5) \right]
	\label{eq:LmunuApp}
\end{equation}

Expanding the chiral projectors,

\begin{align}
	L_{\mu\nu} =& \frac12 \mathrm{Tr} \left[ (\slashed p_b+m_b) \gamma_\mu (\slashed p_t+m_t) \gamma_\nu \right] 
	\nonumber\\ & - \frac12 \mathrm{Tr} \left[ (\slashed p_b+m_b) \gamma_\mu\gamma_5 (\slashed p_t+m_t) \gamma_\nu \right]
	\nonumber\\ & - \frac12 \mathrm{Tr} \left[ (\slashed p_b+m_b) \gamma_\mu (\slashed p_t+m_t) \gamma_\nu\gamma_5	\right]
	\nonumber\\ & + \frac12 \mathrm{Tr} \left[ (\slashed p_b+m_b) \gamma_\mu\gamma_5 (\slashed p_t+m_t) \gamma_\nu\gamma_5 \right]
	\nonumber\\ \equiv& T_1-T_2-T_3+T_4
	\label{eq:Tdecomp}
\end{align}

\subsection*{Evaluation of $T_1$}

The vector-vector contribution is

\begin{align}
	T_1 &= \frac12 \mathrm{Tr} \left[ (\slashed p_b+m_b) \gamma_\mu (\slashed p_t+m_t) \gamma_\nu \right]
	\nonumber\\ &= 2 \left( p_{b\mu}p_{t\nu} + p_{b\nu}p_{t\mu} - g_{\mu\nu} p_b\!\cdot p_t + 	m_tm_bg_{\mu\nu} \right)
\end{align}

\subsection*{Evaluation of $T_2$}

The mixed axial contribution becomes

\begin{align}
	T_2 &= \frac12 \mathrm{Tr} \left[ (\slashed p_b+m_b) \gamma_\mu\gamma_5 (\slashed p_t+m_t) 	\gamma_\nu \right] 
	\nonumber\\ &= -2i \epsilon_{\mu\nu\alpha\beta} p_b^\alpha p_t^\beta
\end{align}

All terms proportional to one quark mass vanish since the corresponding traces contain an odd number of Dirac matrices.

\subsection*{Evaluation of $T_3$}

Similarly,

\begin{align}
	T_3 &= \frac12 \mathrm{Tr} \left[(\slashed p_b+m_b) \gamma_\mu (\slashed p_t+m_t) \gamma_\nu\gamma_5
	\right]
	\nonumber\\ &= -2i \epsilon_{\mu\nu\alpha\beta} p_b^\alpha p_t^\beta
\end{align}

\subsection*{Evaluation of $T_4$}

The last contribution is

\begin{align}
	T_4 =& \frac12 \mathrm{Tr} \left[ (\slashed p_b+m_b) \gamma_\mu\gamma_5 (\slashed p_t+m_t) 	\gamma_\nu\gamma_5  \right]
\end{align}

Using the anticommutation relation in Eq.\ref{gamma5_anti}, and the cyclic property of the Dirac trace, the trace reduces to

\begin{align}
	T_4 = 2 \left( p_{b\mu}p_{t\nu} + p_{b\nu}p_{t\mu} - g_{\mu\nu} p_b\!\cdot p_t - m_tm_bg_{\mu\nu} 	\right)
\end{align}

Summing all four contributions gives

\begin{equation}
	L_{\mu\nu} = 4 \left( p_{b\mu}p_{t\nu} + p_{b\nu}p_{t\mu} - g_{\mu\nu} p_b\!\cdot p_t + i 	\epsilon_{\mu\nu\alpha\beta} p_b^\alpha p_t^\beta \right)
	\label{eq:Lfinal}
\end{equation}

This expression agrees with an independent symbolic evaluation using the \texttt{FeynCalc} package.

Contracting Eq.\ref{eq:Lfinal} with the polarization sum in Eq.\ref{PolSumApp}, and using the two-body kinematic relations Eq.\ref{kinematicrel}, the contraction becomes

\begin{align}
	L_{\mu\nu}P^{\mu\nu} & = 4 \left[ p_b\!\cdot p_t  +  \frac{ 2 (p_b\!\cdot p_V) (p_t\!\cdot p_V) }{m_V^2}
	\right] \\& = 2 \left[\frac{(m_t^2-m_b^2)^2}{m_V^2} + m_t^2 + m_b^2 - 2m_V^2 \right]
	\label{eq:ContractionFinal}	
\end{align}

Finally,

\begin{equation}
	\overline{|\mathcal A(t\to bV)|^2} 	= G_F^2 |V_{tb}V_{qq'}^\ast|^2 f_V^2m_V^2 \left[		\frac{(m_t^2-m_b^2)^2}{m_V^2} + m_t^2 + m_b^2 - 2m_V^2 \right]
\end{equation}

Therefore, the corresponding kinematic function is

\begin{equation}
	\mathcal K_V = \frac{(m_t^2-m_b^2)^2}{m_V^2} + m_t^2 + m_b^2 - 2m_V^2 
\end{equation}

\section{Derivation of the Helicity Amplitudes}
\label{app:helicity}

In this appendix we derive explicitly the helicity amplitudes used in the main text. Starting from the leptonic tensor

\begin{equation}
	L_{\mu\nu} = 4 \left( p_{b\mu}p_{t\nu} + p_{b\nu}p_{t\mu} - g_{\mu\nu} p_b\!\cdot p_t + i 	\epsilon_{\mu\nu\alpha\beta} p_b^\alpha p_t^\beta \right)
\end{equation}

the helicity amplitude corresponding to the vector-meson polarization $\lambda=0,\pm1$ is defined as

\begin{equation}
	L_\lambda = L_{\mu\nu} \, \varepsilon^\mu(\lambda) \, \varepsilon^{\ast\nu}(\lambda)	\label{eq:LlambdaDef}
\end{equation}

Throughout the calculation we work in the rest frame of the decaying top quark, $p_t^\mu=(m_t,0,0,0)$, where the final-state momenta are $p_b^\mu=(E_b,0,0,-p)$ and $p_V^\mu=(E_V,0,0,p)$ with $E_b= \frac{m_t^2+m_b^2-m_V^2}{2m_t}$, $E_V= \frac{m_t^2-m_b^2+m_V^2}{2m_t}$ and $p= \frac{\sqrt{\lambda(m_t^2,m_b^2,m_V^2)}}{2m_t}$. We have defined $\lambda(a,b,c) = a^2+b^2+c^2- 2ab - 2ac - 2bc$ as Källén function. The polarization vectors are chosen as

\begin{align}
	\varepsilon^\mu(\pm) &= \frac1{\sqrt2} (0,\mp1,-i,0) \\
	\varepsilon^\mu(0) &= \frac1{m_V} (p,0,0,E_V)
	\label{epsilonpm}
\end{align}

For $\lambda=\pm1$,

\begin{equation}
	\varepsilon^\mu(\pm)p_{t\mu}=0, 	\qquad 	\varepsilon^\mu(\pm)p_{b\mu}=0
\end{equation}

Consequently, the first two terms vanish

\begin{equation}
	p_{b\mu}p_{t\nu} \, \varepsilon^\mu(\pm) \varepsilon^{\ast\nu}(\pm) = 0
\end{equation}

\begin{equation}
	p_{b\nu}p_{t\mu} \, \varepsilon^\mu(\pm) \varepsilon^{\ast\nu}(\pm) = 0
\end{equation}

The antisymmetric Levi-Civita term does not vanish for transverse polarization. In the top-quark rest frame, the nonvanishing contribution from the Levi-Civita tensor involves the transverse (1) and (2) components of the polarization vectors. Using Eq. \ref{epsilonpm} one obtains

\begin{equation}
	i\epsilon_{\mu\nu\alpha\beta} p_b^\alpha p_t^\beta \varepsilon^{\ast\mu}(\pm)\varepsilon^\nu(\pm) = 	\mp m_t p
\end{equation}

here the sign follows from the chosen convention for the polarization vectors and $\epsilon_{0123}=+1$.
Finally, using $p_b\!\cdot p_t = m_t E_b$, the transverse helicity amplitudes become
\begin{align}
	L_\pm &= -4 \left[(p_b\!\cdot p_t) \, g_{\mu\nu} \varepsilon^\mu(\pm) \varepsilon^{\ast\nu}(\pm) \right] = 4 \left[\,p_b\!\cdot p_t \mp m_t p \right]
\end{align}

the transverse helicity contractions are therefore

\begin{equation}
	L_\pm =	4m_t(E_b\mp p).
\end{equation}

For the longitudinal polarization, we have

\begin{align}
	p_t\!\cdot\varepsilon(0) &= \frac{m_tp}{m_V}, \\
	p_b\!\cdot\varepsilon(0) &= \frac{p(E_b+E_V)}{m_V}
\end{align}

Substituting these relations into Eq.~(\ref{eq:LlambdaDef}) gives

\begin{align}
	L_0 &= 4 \left[ 2 \frac{m_tp}{m_V} \frac{p(E_b+E_V)}{m_V} + p_b\!\cdot p_t \right]
\end{align}

Using the kinematic relations

\begin{equation}
	E_b=\frac{m_t^2+m_b^2-m_V^2}{2m_t}, \qquad p^2= \frac{\lambda(m_t^2,m_b^2,m_V^2)}{4m_t^2}
\end{equation} 

the longitudinal helicity contraction can be expressed entirely in terms of the particle masses as

\begin{equation}
	L_0 = \frac{2}{m_V^2} \left[ (m_t^2-m_b^2)^2 - m_V^2(m_t^2+m_b^2) \right]
\end{equation}

\end{document}